\documentclass[journal,article,submit,moreauthors,pdftex]{mdpi} 
\graphicspath{{./fig/}}

\firstpage{1} 
\pubvolume{xx}
\issuenum{1}
\articlenumber{5}
\pubyear{2020}
\copyrightyear{2020}
\history{Received: date; Accepted: date; Published: date}

\usepackage{amsthm}
\usepackage{amssymb}
\DeclareUnicodeCharacter{00A0}{ }

\Title{Phase transition in Modified Newtonian Dynamics (MONDian) self-gravitating systems}

\Author{Mohammad H. Zhoolideh Haghighi $^{1,2}$*\orcidA{}, S. Rahvar $^{3}$ and M Reza Rahimi Tabar $^{3,4,}$}

\AuthorNames{Firstname Lastname, Firstname Lastname and Firstname Lastname}

\address{%
	$^{1}$ \quad Department of Physics, K.N. Toosi University of Technology, P.O. Box 15875-4416, Tehran, Iran\\
	$^{2}$ \quad School of Astronomy, Institute for Research in Fundamental Sciences (IPM), Tehran, 19395-5746, Iran; \\
	$^{3}$ \quad Department of Physics, Sharif University of Technology, P.O.
	Box 11155-9161, Tehran, Iran\\
	$^{4}$ \quad Institute of Physics and ForWind, Carl von Ossietzky University of Oldenburg, Carl-von-Ossietzky-Stra\ss{}e~9--11, 26111~Oldenburg, Germany.}

\corres{Correspondence: mzhoolideh@ipm.ir}

\abstract{ We study the statistical mechanics of binary systems under gravitational interaction of the Modified Newtonian Dynamics (MOND) in three-dimensional space. Considering the binary systems, in the microcanonical and canonical ensembles, we show that in the microcanonical systems, unlike the Newtonian gravity, there is a sharp phase transition, 
with a high-temperature homogeneous phase and a low temperature clumped binary one.
Defining an order parameter in the canonical systems, we find a smoother phase transition and identify the corresponding critical temperature in terms of physical parameters of the binary system.}


\DeclareFontFamily{OT1}{pzc}{}
\DeclareFontShape{OT1}{pzc}{m}{it}%
{<-> s * [1.25] pzcmi7t}{}
\DeclareMathAlphabet{\mathpzc}{OT1}{pzc}%
{m}{it}

\begin{document}
	
\section{Introduction}
The growth of structures from the initial condition in the early Universe to the galaxies and clusters of galaxies is addressed by the standard model of cosmology. In this scenario, the origin of the structures is quantum fluctuations of a scalar field, the so-called inflaton field. The amplitude of the structures grows after the end of inflation. 
The standard paradigm for structure formation, $\Lambda$CDM \cite{Blumenthal}, has a very good agreement in the early Universe from the CMB observations  \cite{Planck} because the free parameters are tuned to fit the CMB. However, $\Lambda$CDM runs into severe difficulties with local and large scale structures \cite{2021MNRAS.500.5249A,2020MNRAS.499.2845H,MNRAS.491.3042P,2021ApJ...908L..51S}. 

There is another approach to structure formation theory from the statistical mechanics point of view, where the structures form when a phase transition occurs in gravitating systems \cite{Emden1, deVega1, Chavanis, Rahimi}. One classical example in statistical mechanics is the 2D self-gravitating system with a logarithmic gravitational potential \cite{Rahimi}. In this approach, one takes an ensemble of $N$-body particles as a thermodynamical system, where from the partition function one can derive the thermodynamical quantities. We note that in this system all the particles are under their mutual gravitational interaction; this has to be taken into account in calculating the partition function. It has been shown that for such a system there exist two phases of (a) high-temperature gaseous phase, and (b) clumped low-temperature phase. This approach has an analytical solution only in 2D (logarithmic) gravity. To extend it to 3D, one has to deal with a simpler thermodynamical approach such as studying two-body systems \cite{Emden1}. 

In the standard theory of structure formation, to have a compatible theoretical result with observation, we need to have a dark matter component in addition to the baryonic component of the cosmic fluid. The effect of dark matter in structure formation is important when the universe was dominated by radiation whose pressure repelled the baryonic matter and prevented the formation of baryonic structures. The perturbation in the dark matter fluid of the Universe (unlike the baryonic and radiation components) would grow through the gravitational instability to form over-dense regions. The consequence of dark matter structures is that after recombination, the gravitational potential of dark matter could accumulate the baryonic matter to form the baryonic part of galaxies within dark matter halos \cite{Peebles,1993Padmanabhan}.
There is another approach to deal with the dynamics of large scale structures by replacing the dark matter with a modification to the gravity law. In some of these models, the extra degrees of freedom such as the scalar or the vector sectors are considered for the gravitational field. A generic formalism of this theory is the Scalar-Vector-Tensor theory which is the so-called Modified gravity model (MOG) \cite{moffat}. Although this theory can explain the dynamics of galaxies and clusters of galaxies without a need for dark matter \cite{rahvar1,rahvar2}, it predicts unexpected mass-to-light ratio \cite{2019ApJ...884L..25H}. There are also non-local gravity models, where in one of these theories, the Einstein gravity has been supplemented by the non-local terms, in analogy to the non-local 
electrodynamics \cite{mashhoon}. This theory also educes to the standard Poisson equation in the weak field approximation with an extra term that plays the role of dark matter. The theory provides compatible dynamics for galaxies and clusters of galaxies without the need for a dark matter component \cite{rahvar3}. There is also another popular model which is called Modified Newtonian Dynamics (MOND) \cite{milgrom}. In this theory, Newton's second law or usually the Poisson equation is modified in systems with accelerations smaller than a universal acceleration, $a_0$. This theory also provides compatible dynamics to the spiral and elliptical galaxies \cite{McGaugh}. It is worth mentioning that the main problem of modified gravity models is that they can not explain the observational data on different scales, though see \cite{2020MNRAS.499.2845H} for a possible hybrid solution using both MOND and hot dark matter in the form of sterile neutrinos.

In this work, we study the phase transition for a binary interacting via MONDian gravity, in  non-expanding and expanding spaces. For simplicity, we start with an ensemble of two-body objects instead of the $N$-body system and investigate the phase transition by decreasing the temperature of the system. 

The rest of the paper is organized as follows: In section II, we review the statistical mechanics of binary interacting systems under a Newtonian potential in the microcanonical ensemble. In section III, we present the statistical mechanics of a binary system
in MOND. Section IV is devoted to studying the statistical mechanics of MONDian systems in the canonical ensemble. We then study and discuss the influence of expanding Universe with scale factor $a(\tau)$ on the critical temperature of the phase transition in section V.  Section VI summarizes the paper and discusses
possible further research.

\section{Statistical mechanics of a self-gravitating binary under Newtonian potential: Microcanonical ensemble}
\label{sec2}

The statistical mechanics of self-gravitating systems has been the subject of attention for many years \cite{Emden1, deVega1, Levin1}.  Such systems have distinguished physical properties due to the long-range nature of the gravitational force. At thermal equilibrium, these systems are not spatially homogeneous, and intrinsic inhomogeneity character suggests that fractal structures can emerge in a system of gravitationally interacting particles \cite{deVega1}. Here we review the statistical mechanics of a self-gravitating binary system in Newtonian gravity in the microcanonical ensemble \cite{Emden1}. We start with the Hamiltonian of a two-body system,

\begin{equation}
H ( {\bf P}, {\bf Q}; {\bf p}, {\bf r}) = {{\bf P}^2 \over 2M} + {{\bf p}^2 \over 2 \mu}+ V(r) \quad .
\label{hamil}
\end{equation}
where in 3D the potential is $V(r) = - Gm^2/r$,  $( {\bf Q}, {\bf P}) $ are coordinates and momentum of the center of mass, and 
$({\bf r}, {\bf p})$ are the relative coordinates and momentum with the reduced mass. In what follows, for the sake of simplicity we take the mass of these two bodies to be identical,
(i.e. $M = 2 m$ and $\mu = m/2$).
We assume a spherical shape for the two objects with radius of ${b}/{2}$ and that the two-body system is confined in a spherical box of radius $R$, where $r$ in equation (\ref{hamil}) varies within the interval of $(b,R)$. The volume associated with a constant energy of this system (i.e. $H=E$) in the phase-space \cite{2008arXiv0812.2610P,2002LNP...602.....D} is given by the density of states
\begin{equation}
g(E)= \int \delta (E-H(r,p,Q,P)) d^3P d^3p d^3Q d^3r \quad .
\label{pv}
\end{equation}
First we integrate over $Q$, which leads to $4{\pi}r^3/3$, and inserting the explicit form of  the Hamiltonian leads to the following relation:
\begin{equation}
g(E)= \frac{4 \pi R^3}{3}\int d^3P d^3r \int_0^{\infty}  \delta (E-{{\bf p}^2 \over 2 \mu} - {{\bf P}^2 \over 2M} - V(r)) 4 \pi p^2 dp  \quad.
\label{eq2}
\end{equation} 
Now we perform integration over $p$-space. Using the new variable $x\equiv p^2/\mu$, the square-root term appears from the integration as a result of a property of the Dirac $\delta$-function. 
\begin{equation}
g(E)= \frac{8 \pi^2 m^{3/2}R^3}{3}\int d^3P d^3r  \sqrt{E-\frac{P^2}{2m} - V(r)}  \quad. 
\end{equation} 
This time we integrate over $P$-space similar to the previous step in equation (\ref{eq2}). The result is

\begin{equation}
g(E)= AR^3\int_b^{r_{\rm max}} r^2 dr \left(E-V(r)\right)^2 \quad ,
\label{pv1}
\end{equation} 
where $A= 64 \pi^5 m^3/3$. In the case of Newtonian gravity, the potential energy is $V(r)=- Gm^2/r$
and since the kinetic energy is always positive definite (i.e $E-V(r)>0$), the upper limit of the integral ($r_{max}$) should be taken in such a way that guarantees the positive sign of the kinetic energy. The upper bound of integration for the following ranges of energy is given by,
\begin{equation}
\Bigg \{
\begin{tabular}{c}  
$(-Gm^2/b)<E<(-Gm^2/R) \hskip 1cm r_{max}=Gm^2/E$  \\
$(-Gm^2/R)<E<+ \infty \hskip 1.5cm r_{max}=R $
\end{tabular}
\end{equation}
The second condition has no specific meaning in astrophysics, though it has an analogy with the standard thermodynamics where the size of the box is $R$. We assume that the container has a fixed volume while we can increase the velocity of the particles, and they are constrained to stay in this volume. 
Integrating (\ref{pv1}) results in 

\begin{equation}
\frac{g(E)}{A(Gm^2)^3}=
\left\{ 
\begin{array}{l}
{R^3\over3}(-E)^{-1} \left( 1+{bE\over Gm^2}\right)^3 ,
\hspace{1cm}(-Gm^2/b)< E<(-Gm^2/R)\\
{  }\\
{R^3\over3} (-E)^{-1} \left[ \left(1+{RE\over Gm^2}\right)^3
- \left(1+{bE\over Gm^2}\right)^3 \right],
\quad (-Gm^2/R )<E< \infty  .
\end{array}
\right.
\label{gee}
\end{equation}

\begin{figure}
	\begin{center}
		\includegraphics[width = 7cm]{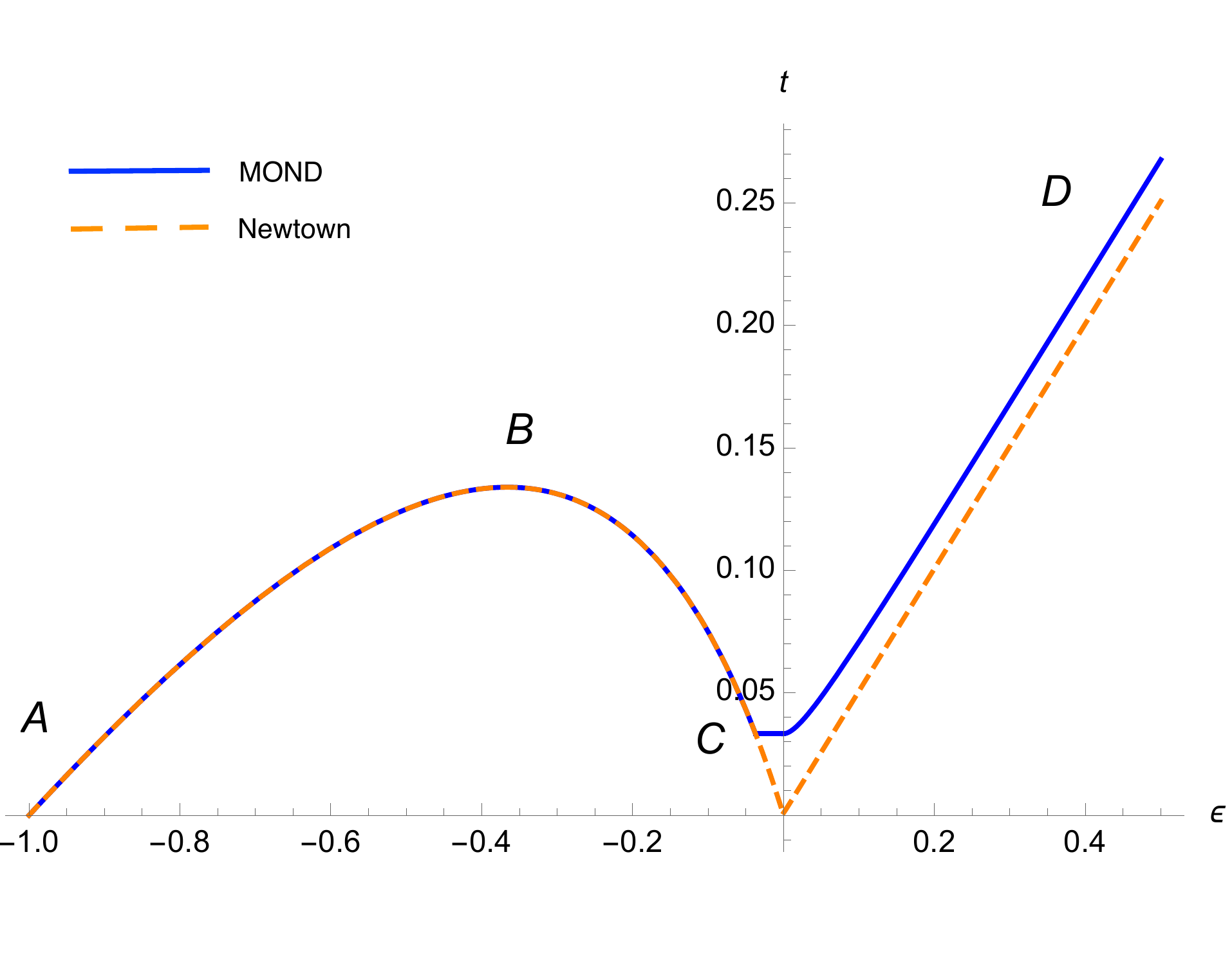}
		\caption{Dimensionless temperature versus dimensionless energy (Eq. \ref{temp}) for a binary system interacting under Newtonian gravity (dashed orange curve) and MONDian gravity (solid blue curve). As expected, there is a flat part for MOND due to entering the DML. Since for constant temperature the energy has a finite change there must be a sharp phase transition.}
		\label{fig1}
	\end{center}
\end{figure}

From the density of states function $g(E)$ which represents the  phase space volume covered by this system with energy $E$, we can calculate the entropy and the temperature of the system according to the following relations (with Boltzmann constant $K_B=1$):
\begin{equation} S(E)=\ln g(E);\quad T^{-1}(E)=\beta (E)= {\partial S(E)\over \partial E} \quad .
\label{entro}
\end{equation} 
From equation (\ref{gee}), we obtain the dimensionless temperature for the interval 
$(-Gm^2/b)< E<(-Gm^2/R)$ as
\begin{equation} t(\epsilon)= \left[{3\over 1+\epsilon}
-{1\over \epsilon}\right]^{-1} \quad .\label{temp}
\end{equation}
where the dimensionless temperature  $t$ and energy $\epsilon$ are defined as 
\begin{equation}
t\equiv (bT/Gm^2), \hskip 1cm \epsilon \equiv (bE/Gm^2) \quad .
\label{t}
\end{equation}
 Also for the second interval of $-Gm^2/R <E< \infty$ in (\ref{gee}), 
$t(\epsilon)$ is given by 
\begin{equation} t(\epsilon)= \left[ { 3\left[ (1+\epsilon)^2- {R\over b}(1+ {R\over b}\epsilon)^2\right]
	\over (1+\epsilon)^3 -(1+{R\over b}\epsilon)^3}
-{1\over \epsilon} \right]^{-1}  \quad .
\label{qtemp}
\end{equation}
In Figure (\ref{fig1}), we depict $t(\epsilon)$ in terms of $\epsilon$ \cite{2008arXiv0812.2610P}. It can be seen that the specific heat is positive along AB and CD while it is negative along BC. For a system with energy  in the range AB, the two solid spherical objects are in contact, and increasing the energy of the system increases the kinetic energy, or in other words, the temperature of the system. Over the range BC, the two objects detach from each other and start circular motion according to the mutual gravitational force between them. 
For the CD path in Figure (\ref{fig1}) the total energy is larger than zero (i.e. $E>0$) and the two objects decouple from each other and behave as free particles. 

\section{Statistical mechanics of a self-gravitating binary in MOND: Microcanonical ensemble }
\label{sec3}
In the Modified Newtonian Dynamics (MOND), proposed to solve the dark matter problem on galaxy scales\cite{milgrom}, the second law of Newtonian mechanics is modified for small accelerations. The characteristic acceleration for this modification is  $a_{0}=1.2 \times 10^{-10} m s^{-2}$ \cite{1991MNRAS.249..523B} where for $a\leq a_0$, the definition of force is modified to $F = ma\mu(a/a_0)$. Here $\mu(x)$ is larger than unity for smaller accelerations and is unity for the large accelerations (i.e. $a\gg a_0$). This model can explain the rotation curve of spiral galaxies and predicts the Tully-Fisher relation \cite{Tully}. However, in this theory, energy and momentum are not well-defined \cite{Felten}. One of the simple solutions to this problem is that one may interpret MOND as modified gravity rather than modified dynamics. The gravitational acceleration   in the deep MOND limit where $ g\ll a_{0}$ will be  \cite{milgrom}
\begin{equation}
g_{DML}=-\sqrt {g_{N}.a_{0}}=-\frac{\sqrt{Gma_0}}{r} \quad .
\end{equation}
The subscript DML stands for the deep-MOND limit, in which the potential from the gravitational acceleration is given by 
\begin{equation}
\phi_{DML} = \sqrt{Gm a_{0}}~\ln(r/R) \quad.
\end{equation} 
For an isolated binary system in the DML, the gravitational force between the two particles scales as $F \sim 1/r$ which for  objects in circular motion results in a constant velocity, compatible with the flat rotational curves of galaxies. We note that for $g \gg a_0$ the Newtonian gravity is recovered (i.e. $g=g_{N}$) and the potential will behave as $\phi_N = -Gm/r$. 

Since in MOND the acceleration of a test particle in a gravitating system is stronger than in Newtonian gravity, in addition to studying the dynamics of a system, we can investigate the formation 
of structures in these two scenarios.  For a self-gravitating system, the free-fall time scale represents the strength of clustering of a structure. The ratio of characteristic timescales in Newtonian and MOND gravity is \cite{1987gady.book.....B, Golovnev},
\[
\frac{T_{MOND}}{T_N} = \Big[ \frac{16}{\pi^2} \cdot  \frac{g_N}{a_0} \Big]^{\frac{1}{4}} \quad.
\]
For $g_N<a_0$ we would expect a shorter time-scale for the clustering in MOND compare to the Newtonian gravity. We refer interested readers to \cite{2021MNRAS.506.5468Z} for a thorough investigation into the free-fall timescales in MOND.

In order to have a continuous transition from the DML to the Newtonian gravity, one needs a transition function $\mu(a)$ \cite{Beyken and milgrom}. There are various transition functions. One of the simple models is $\mu(x) = x/(1+x)$ or a sigmoid function,  other transition function can be found in  \cite{Famaey, Banik}. The sigmoid function better fits the observational data \cite{2018MNRAS.480.2660B}. 

Now let us study the statistical mechanics of two-body objects in MOND in the microcanonical ensemble.
The enumeration of the total number of possible states (phase volume) for a binary system with energy $E$ is given by 
\begin{equation}
g(E)= \int \delta (E-H(r,p,Q,P)) d^3P d^3p d^3Q d^3r \quad .\label{pv}
\end{equation}
By integrating over $P$, $p$ and $Q $, ( similar to the calculations in section (2) for Newtonian gravity, the function $g(E)$ is given by
\begin{equation}
g(E)= AR^3\int_b^{r_{\rm max}} r^2 dr \left(E-V(r)\right)^2 \quad ,
\label{pv}
\end{equation} 
where $A= 64 \pi^4 m^3/3$. Taking into account the Newtonian and DML phases for the gravitational potential, the total potential is given in the following two domains,
	\begin{equation}
V(r)=\Bigg \{
\begin{tabular}{c}  
$-Gm^2/r,\hskip 2.5cm  r<r_{M}$  \\
\\
$	m\sqrt{Gm a_{0}}ln(r/R), \hskip 1cm r>r_{M}$
\end{tabular}
\end{equation}
Where the MOND radius $r_{M}=\sqrt{Gm/ a_{0}}$ is the scale of transition between the two domains. In order to calculate integral (\ref{pv}), one needs precisely determine $r_{max}$ in terms of energy ranges, i.e.,
\begin{equation}
\left\{ 
\begin{array}{l}
\frac{-Gm^2}{b} <E< \frac{-Gm^2}{r_{M}} ,\quad\qquad
r_{max}=\frac{-Gm^2}{E}\\
{  }\\

\frac{-Gm^2}{r_{M}} <E<0  ,\quad\qquad\qquad r_{max}=R \exp(\frac{E}{m\sqrt{Gm a_{0}}})\\
{  }\\

0<E<\infty  ,\quad\qquad\qquad\qquad r_{max}=R \quad 
\end{array}
\right.
\label{gee2}
\end{equation}
Here the bound of $r_{max}$ results from the positive 
sign of the kinetic energy. We define the temperature of system from equation (\ref{entro}) and plot $t=t(\epsilon)$ for three energy ranges in Fig (\ref{fig1}). Here we adopt ${R}/{b}=10^{10}$ and ${e_{2}}/{e_{1}}=100$ where $e_{1}=m\sqrt{Gm a_{0}}$ and $e_{2}=Gm^2/b$. In this figure, the $(t,\epsilon)$ diagram for MONDian gravity is almost similar to that for Newtonian, except for $\epsilon$ just slightly below 0, where for the case of Newtonian gravity $t(0) = 0$ but for MONDian gravity $t(0)>0$. In fact, for MONDian gravity (the solid curve), there is a flat part, in which for constant temperature, the energy has a finite change, indicating a sharp phase transition where the heat capacity diverges. 
 
We can interpret this area as when the binary system enters the deep-MOND where the rotation velocity of the binary objects around their center of mass is $v_{rot} = \sqrt{Gma_0/2}$. In this case, with increasing energy of the system, the orbital size of the binary increases; however, the kinetic energy, which represents the temperature of the system, remains constant. In the next section, we study statistical mechanics and phase transition of a binary system in the MONDian gravity for the canonical ensemble.

\section{Statistical mechanics of a self-gravitating binary in MOND: Canonical ensemble}
\label{sec4}

\begin{figure} 
	\begin{center}
		\includegraphics[scale=0.25]{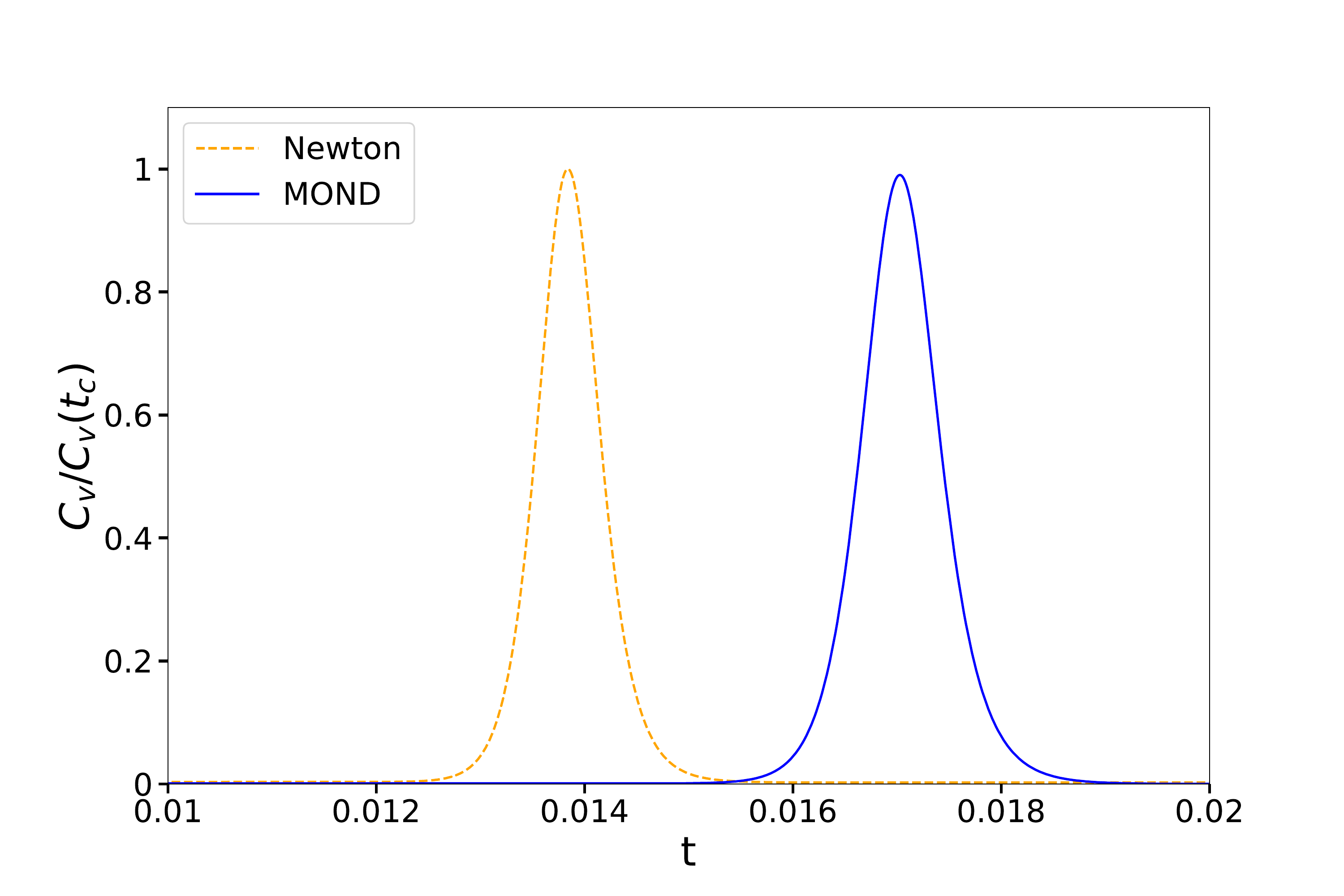}
		\caption{Normalized specific heat of the system made of a binary under Newtonian (dashed orange curve) and MONDian (solid blue curve) gravity. The phase transition temperature is at the peak of these curves. 
			}
		\label{fig3}
	\end{center}
\end{figure}

Let us at first assume an ensemble of binary objects under Newtonian gravity. The system is composed of an ensemble of thermalized binaries with an associated temperature. The partition function associated with a binary system in this ensemble is given by
\begin{equation} 
Z(\beta)=\int d^3Pd^3pd^3Qd^3r \exp (-\beta H) \quad ,
\label{partis1}
\end{equation} 
where the parameters and the Hamiltonian are defined in Sec. (2). Integrating over momenta $P$, $p$ and position $Q$, equation (\ref{partis1}) simplifies to \cite{1987gady.book.....B,2008arXiv0812.2610P}:
\begin{equation} Z(\beta)=R^3 \beta^{-3}\int_b^R dr \, r^2 \exp \left(\beta \frac{GM^2}{r}\right) \quad .
\end{equation}
In dimensionless form, using the definition of $t$ as introduced in Sec. (2), the partition can be written as 

\begin{multline}
Z(t)=\left(\frac{R}{b}\right)^3 t^{3}\int_1^{{R}/{b}} x^2 \exp \left(\frac{1}{t x}\right) \, dx \quad \\
= \frac{R^3}{12 b^6} \Bigg[ b^3 (-2 \text{Ei}\left(\frac{b}{R t}\right)-\log \left(\frac{R t}{b}\right)+2 \log \left(-\frac{R t}{b}\right)+\log \left(\frac{b}{R t}\right)+2 \text{Ei}\left(\frac{1}{t}\right)-2 e^{1/t} t \left(2 t^2+t+1\right) \\
 -\log \left(\frac{1}{t}\right)-2 \log (-t)+\log (t))
 +2 R t e^{\frac{b}{R t}} \left(b^2+b R t+2 R^2 t^2\right)\Bigg]
 \label{Ei}
\end{multline}
 Equation (\ref{Ei}) is expressed in term of exponential integral function $\left(Ei(x)= -\int_{-x}^{\infty} \frac{\exp \left(-t \right)}{t} \,dt  \right)$ and has no known analytical solution, so here we solve it numerically and calculate the mean energy of the system by using:
\begin{equation}
E(\beta)= -\partial \ln Z/ \partial\beta \quad.
\label{E}
\end{equation}
In order to identify a phase transition, we calculate numerically the derivative of energy with respect to the thermodynamical variables. If this quantity diverges or becomes discontinuous, the system undergoes a phase transition \cite{2011ConPh..52...98S}. We calculate the derivative of energy with respect to the temperature, which is defined as the specific heat and is shown in Fig. (\ref{fig3}).

\begin{figure} 
	\begin{center}
		\includegraphics[scale=0.25]{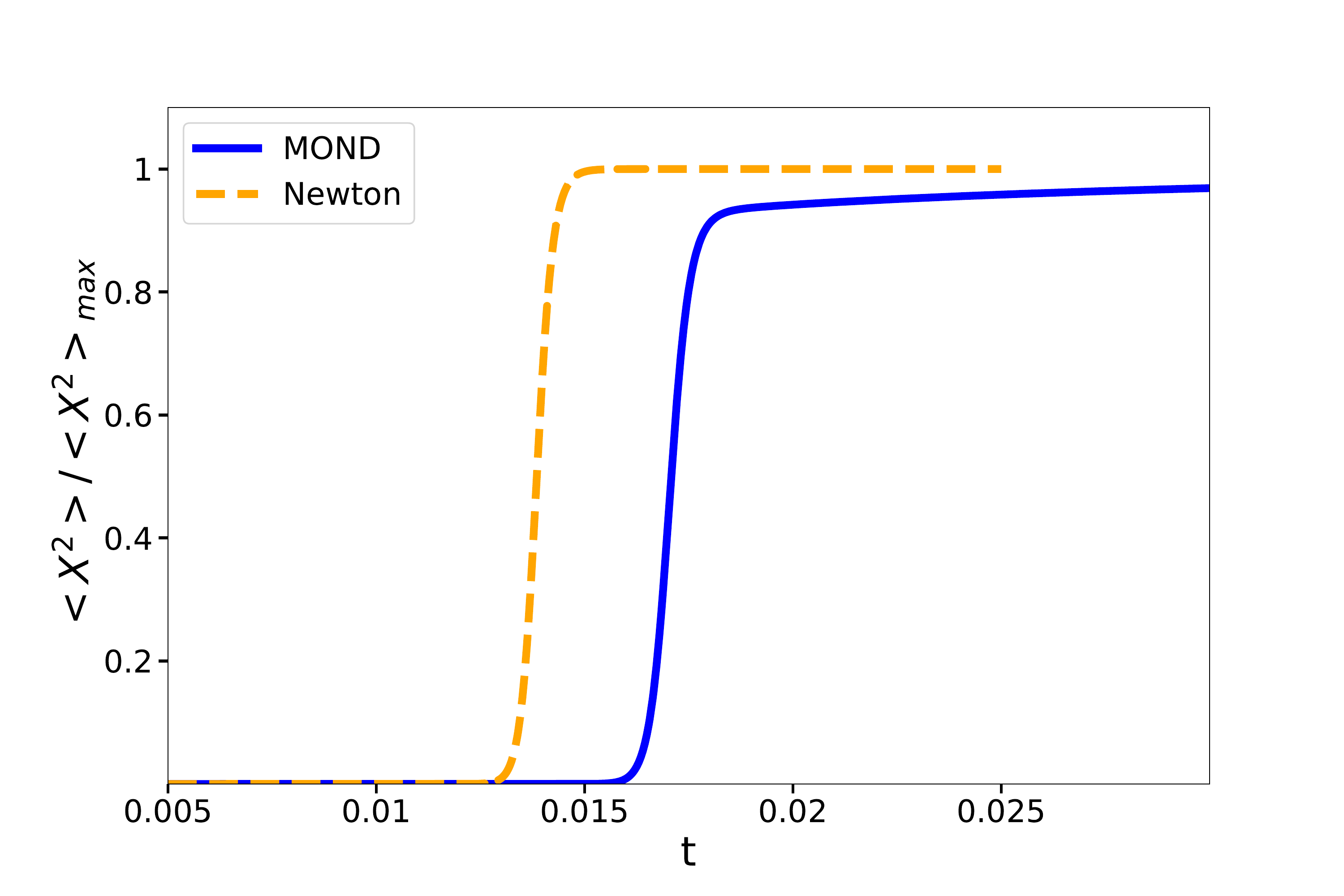}
		\caption{Order parameter as a criterion to detect the phase transition. The orange dashed curve represents the Newtonian system, and the blue solid one is for MOND. This plot shows that the mean value of the distance between the particles changes very fast but remains differentiable near the phase transition. 
		}
		\label{fig4}
	\end{center}
\end{figure}

From the numerical calculation of specific heat $c_v$ in terms of the temperature, we find that it has a peak at $t = t_{critical}$.
To understand the nature of the detected phase transition, we define an order parameter and study its behavior near the  critical temperature. Here we take the mean distance between the companions of a binary system as the order parameter and define it as \cite{Rahimi}: 
\begin{equation}
<r^2>=\frac{\int_b^R dr r^4 \exp (\beta \frac{Gm^2}{r})}{\int_b^R dr r^2 \exp (\beta \frac{Gm^2}{r})} \quad.
\label{orderparNew}
\end{equation}
For the first-order phase transition, we expect to have an abrupt change in the order parameter. For instance, if we apply it for matter in the liquid and gaseous phases, this parameter sharply changes when a liquid changes to the gaseous state. For the second order phase transition, the order parameter will be a continuous function at the critical temperature. 
For simplicity, we rewrite Eq.(\ref{orderparNew}) in dimensionless form as:
\begin{equation}
<x^2>=\frac{\int_1^{{R}/{b}} x^4 \exp \left(\frac{1}{t x}\right) \, dx}{\int_1^{{R}/{b}} x^2 \exp \left(\frac{1}{t x}\right) \, dx} \quad 
\end{equation} 
and calculate $<x^2>$ numerically allowing us to plot the order parameter as a function of temperature in Fig. (\ref{fig4}).
We notice that there is the phase transition for the order parameter at exactly the same temperature that the specific heat has a peak, whereas the order parameter is a differentiable function at the critical temperature. The order parameter shows that a high-temperature system has a homogeneous phase, while at low temperature $<x^2>$ vanishes. 

Now we perform a similar calculation for the temperature dependence of the order parameter in the combination of deep MOND and Newtonian gravity, where for small and large accelerations (i.e $a<a_0$ and $a>a_0$), DML and Newtonian gravity will have dominant contributions, respectively. 
Since the potential is a function of distance, we need to make an approximation before calculating the partition function. The potential is defined as $\phi(r) = -\int F.dr$, we can break this integration into separate three parts with different gravitational potentials. For $r\ll r_{M}$ we have Newtonian gravity, and $r\gg r_{M}$ potential is in DML, and between we are in a regime that is a combination of Newtonian and MONDian, i.e.,
\begin{eqnarray}
\phi(r) &=& -\int F.dr = 
-\int _{0<r<r_{M}-\delta} (-GM/r^2)dr  \cr \nonumber \\
&-&\int_{r>r_{M}+\delta} (-\sqrt{(GMa_0/r^2)})dr 
-\int_{r_{M}-\delta<r<r_{M}+\delta} F.dr 
  \quad .
\end{eqnarray}
Three terms in the integration belong to the Newtonian, DML, and mixing of Newtonian and DML, respectively. Here $\delta$ is a small constant. The result of integration is,
\begin{equation}
\phi(r) = -GM/r + \sqrt{GMa_0}\ln(r) -\int_{ r_{M} -\delta}^{ r_{M} +\delta} F.dr+ C \quad .
\end{equation}
where C is the constant of integration, and we chose it to be $\sim -\ln(R)$.
We note that the third term in r.h.s of Eq.(26), that is $\int_{ r_{M} -\delta}^{ r_{M} +\delta} F.dr$, depends on the interpolating function, ($F=m\mu(a/a_0)$). A simple approach is by choosing a proper $\mu$ with a very fast transition from Newtonian regime to MONDian regime. Then we can ignore this term and the potential simplifies to 
\begin{equation}
\phi(r) \simeq -GM/r |_{(r<r_{M})}- \sqrt{(GMa_0)}\ln(r/R)|_{(r>r_{M})} \quad.
\label{Mond1}
\end{equation}

An alternative approach is to use a simple interpolating function for a single point mass using a hyperbolic substitution \cite{Famaey, Banik},
\begin{eqnarray}
	\phi \left( r \right) &=& \sqrt{Gma_{_0}} \left[ \ln \left(1 + \sqrt{1 + \tilde{r}^2} \right) - \frac{1}{\tilde{r}}
	 - \sqrt{\frac{1}{\tilde{r}^2} + 1}  \right] \quad . \nonumber \\
	~~{where} \quad  \tilde{r} &\equiv& \frac{2r}{r_{_M}} \quad and  \quad  r_{_M} \equiv \sqrt{\frac{Gm}{a_{_0}}} \quad .
  \label{Potential}
\end{eqnarray}
We provide the influence of the interpolating function on the critical temperature, see below.

For simplification, we continue with Eq. \ref{Mond1}.  As a result, the partition function of the ensemble of the binary system in the MONDian gravity (using the MONDian potential in Eq.~\ref{partis1}) by integrating over variables $P,p$ and $Q$ is
\begin{equation} Z(\beta)=R^3 \beta^{-3}\Big[\int_b^{r_{M}} dr r^2 \exp(\beta \frac{Gm^2}{r}) + \int_{r_{M}}^R dr r^2 \exp ( -e_{1}\beta ln({r}/{R})) \Big] \quad ,
\label{partition1}
\end{equation}
Which in dimensionless representation it simplifies to:
\begin{equation}
Z(t)=(\frac{R}{b})^3 t^{3}\Big[\int_1^{{r_{M}}/{b}}dx x^2 \exp( \frac{1}{t x}) + \int_{{r_{M}}/{b}}^{{R}/{b}}dx x^2 \exp( -\frac{e_{1} \ln(bx/R)}{e_{2} t}) \Big] \, \quad. 
\end{equation}
Here $e_{1}=m\sqrt{Gm a_{0}}$ and $e_{2}=Gm^2/b$.
Following the same procedure as the Newtonian case, we calculate the specific heat and order parameter and plot them in Figs. \ref{fig3} \& \ref{fig4}, where the specific heat and the order parameter are given by:
\begin{equation}
C_{v}=\partial E / \partial T=\frac{ \partial (t^2 \partial\ln Z/ \partial t)}{ \partial t } \quad ,
\end{equation}
\begin{equation}
<r^2>=\frac{\int_b^{r_{M}} r^4 \exp (\beta {Gm^2}/{r})dr+\int_{r_{M}}^R r^4 \exp (-\beta e_{1}ln(r/R))dr}{\int_b^{r_{M}} r^2 \exp (\beta {Gm^2}/{r})dr+\int_{r_{M}}^R r^2 \exp (-\beta e_{1}ln(r/R))dr} \quad .
\label{orderpar}
\end{equation}
Eq. \ref{orderpar} can also be rewritten in the dimensionless form as:
\begin{equation}
<x^2>=\frac{\int_1^{\frac{{r_{M}}}{b}} x^4 \exp \left(\frac{1}{t x}\right) \, dx+\int_{\frac{{r_{M}}}{b}}^{\frac{R}{b}} x^4 \exp \left(-\frac{{e_{1}} \ln (b x/R)}{{e_{2}} t}\right) \, dx}{\int_1^{\frac{{r_{M}}}{b}} x^2 \exp \left(\frac{1}{t x}\right) \, dx+\int_{\frac{{r_{M}}}{b}}^{\frac{R}{b}} x^2 \exp \left(-\frac{{e_1} \ln (b x/R)}{{e_2} t}\right) \, dx} \quad .
\end{equation}

Comparing Figs. \ref{fig3} and \ref{fig4}, the phase transition temperature is identical whether we obtain it from the divergence of $C_{v}$ or from a fast change of $<x^2>$. Also, we note that $C_v$ is always positive for the canonical case, unlike the micro-canonical case.

\begin{figure} 
	\begin{center}
		\includegraphics[scale=0.25]{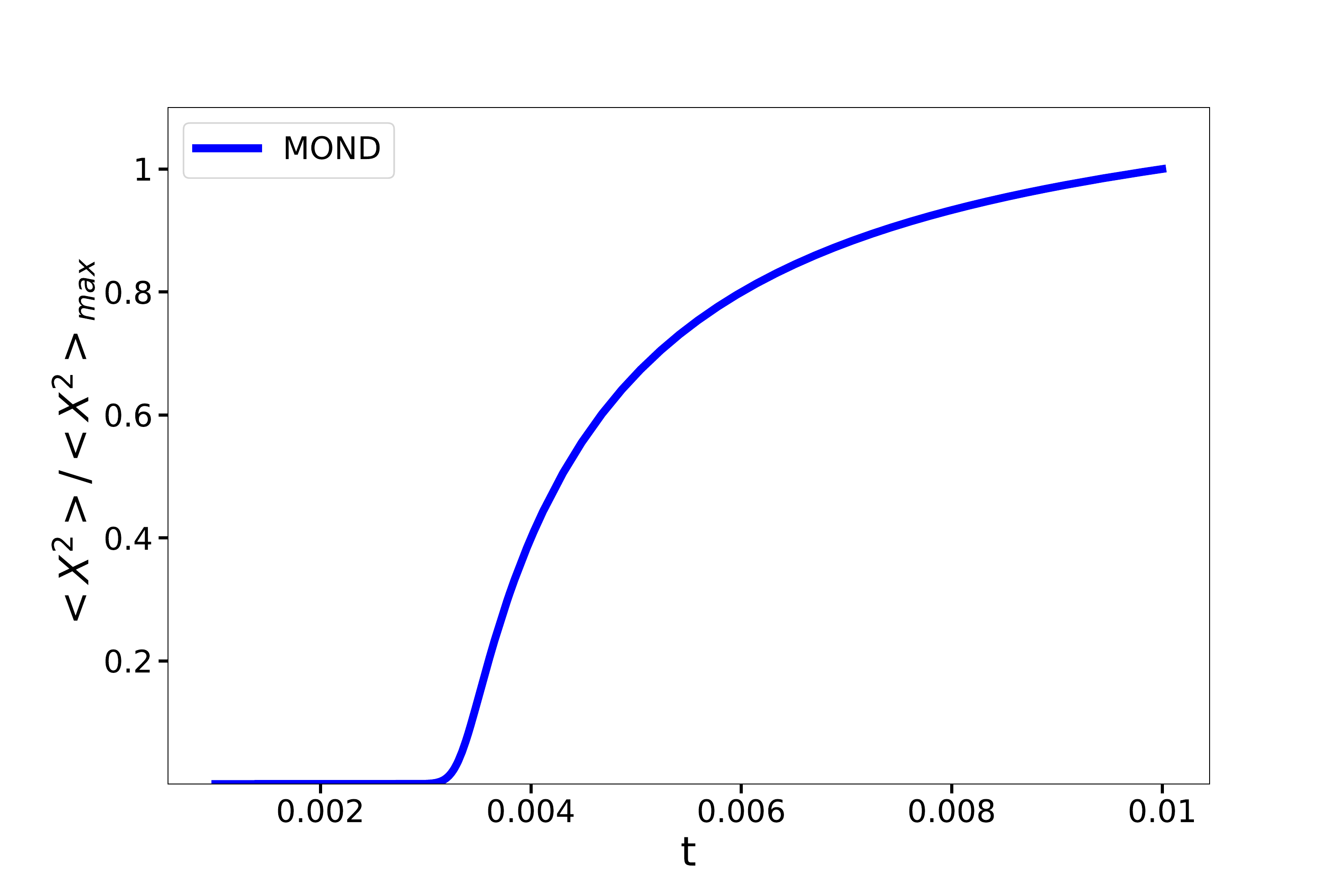}
		\caption{Order parameter as a criterion to detect the phase transition. This plot shows that the variance  of the distance between the particles changes fast but remains differentiable near the phase transition. 
		}
		\label{fig:interpol}
	\end{center}
\end{figure}

If we use the specific interpolating function that is introduced in Eq. (\ref{Potential}), we observe a similar behavior, but at a different critical temperature (see Fig \ref{fig:interpol}). We see that qualitatively the critical behavior of systems is the same but the interpolating function affects the details of the phase transition.

\section{Statistical mechanics of a self-gravitating binary in MOND: Comoving coordinates}
In an expanding universe with characteristic scale factor $a(\tau)$, the physical coordinates of  $r$ is related to the comoving coordinate $q$ as \cite{deVega2},
\begin{equation}
    r \equiv a(\tau) q \quad.
\end{equation}
The Hamiltonian of a binary system with potential given in Eq. (\ref{Mond1}) can be written in the comoving coordinate  as
\begin{equation}
    H = \frac{p^2}{2m {a(\tau)}^2} - \frac{Gm^2 }{a(\tau) q} \left|_{(q<q_{M}{a(\tau)})} + {m}\sqrt{Gma_0}\ln(qa(\tau)/R)\right|_{(q>q_{M} {a(\tau)})}  \quad .
    \label{comoving:hamilton2}
\end{equation}
This Hamiltonian in an expanding universe can also be obtained from the Minkowski-Hamiltonian with the following replacements:
\begin{equation}
    m \rightarrow ma(\tau)^2, \quad \quad G \rightarrow G{a(\tau)}^{-5}, \quad \quad  a_0 \rightarrow a_0 {a(\tau)}^{-1}, \quad \quad  R \rightarrow R {a(\tau)}^{-1} \quad .
\end{equation}
We can interpret the renormalization process by assigning a dynamics to these parameters. For instance the $ m \rightarrow ma(\tau)^2$ renormalization is related to the kinetic energy of particles that decrease as $1/a^2$ with the expansion of the Universe.  Here we have the effective 
gravitational constant $(G \rightarrow G{a(\tau)}^{-5})$ as well as the acceleration parameter of MOND ($a_0 \rightarrow a_0 {a(\tau)}^{-1}$) changing with the scale factor.   \\
To consider a self-gravitating binary at any time in approximate thermal equilibrium, we assume that the characteristic time of the particle motions under their mutual gravitation is shorter than the time variation of the scale factor. This hypothesis is valid for structures that are almost decoupled from the expansion and become virialized. 
Given this, the Eqs. \ref{partition1} \& \ref{orderpar} in the comoving frame will be as follow:
\begin{equation}
   Z(\beta) = R^3 a(\tau)^3 \beta^{-3}  \Bigg[\int_{b}^{q_{M}} q^2  \exp (\beta {Gm^2}/{a(\tau)q})dq+\int_{q_{M}}^{R} q^2  \exp (-\beta m \sqrt{Gma_0}ln(a(\tau)q/R))dq\Bigg] \quad .
\label{comov:parti1}
\end{equation}
\begin{equation}
<q^2>= a(\tau)^{2} \frac{\int_{b}^{q_{M}} q^4  \exp (\beta {Gm^2}/{a(\tau)q})dq+\int_{q_{M}}^{R} q^4  \exp (-\beta m \sqrt{Gma_0}ln(a(\tau) q/R))dq}{\int_{b}^{q_{M}} q^2  \exp (\beta {Gm^2}/{a(\tau)q})dq+\int_{q_{M}}^{R} q^2  \exp (-\beta m \sqrt{Gma_0}ln(a(\tau)q/R))dq} \quad .
\label{comov:orderpar}
\end{equation}
Eqs. \ref{comov:parti1} \& \ref{comov:orderpar} can be rewritten in dimensionless form as:
\begin{equation}
   Z(t) = (\frac{R}{b})^3 a(\tau)^{3} t^3 \Bigg[\int_1^{\frac{{q_{M}}}{b}} x^2 \exp \left(\frac{1}{t x a(\tau)}\right) \, dx+\int_{\frac{{q_{M}}}{b}}^{\frac{R}{b}} x^2 \exp \left(-\frac{{e_1} \ln (b a(\tau) x/R }{{e_2}  t}\right) \, dx\Bigg] \quad ,
  \label{comov:parti2}
\end{equation}
\begin{equation}
<x^2>= a(\tau)^{2} \frac{\int_1^{\frac{{q_{M}}}{b}} x^4 \exp \left(\frac{1}{t x a(\tau)}\right) \, dx+\int_{\frac{{q_{M}}}{b}}^{\frac{R}{b}} x^4 \exp \left(-\frac{{e_{1}} \ln (b a(\tau) x/R)}{{e_{2} } t}\right) \, dx}{\int_1^{\frac{{q_{M}}}{b}} x^2 \exp \left(\frac{1}{t x a(\tau)}\right) \, dx+\int_{\frac{{q_{M}}}{b}}^{\frac{R}{b}} x^2 \exp \left(-\frac{{e_1} \ln (b a(\tau) x/R }{{e_2}  t}\right) \, dx} \quad ,
\label{comov:orderpar2}
\end{equation}
\begin{figure}
\begin{center}
\includegraphics[scale=0.25]{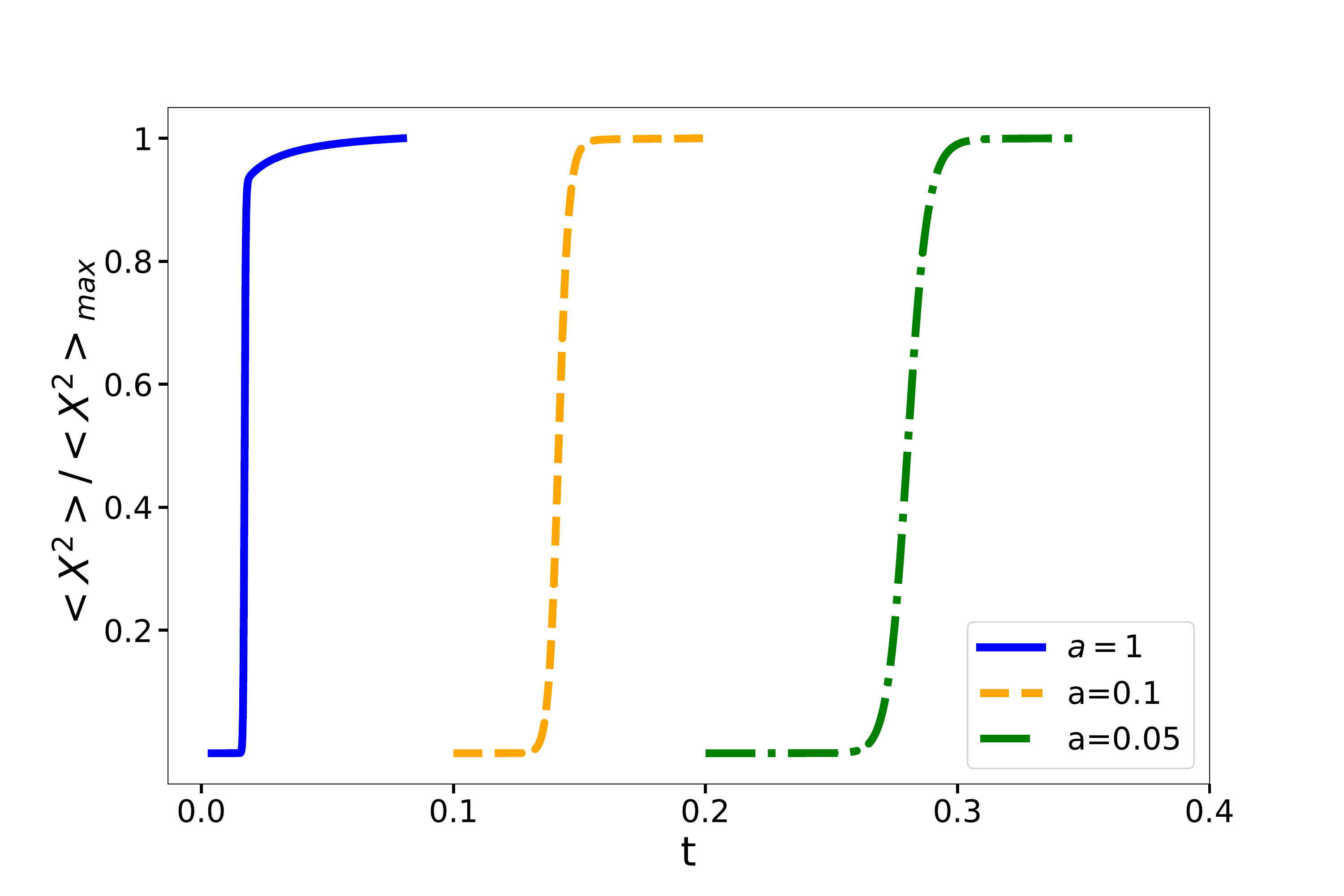}
\includegraphics[scale=0.25]{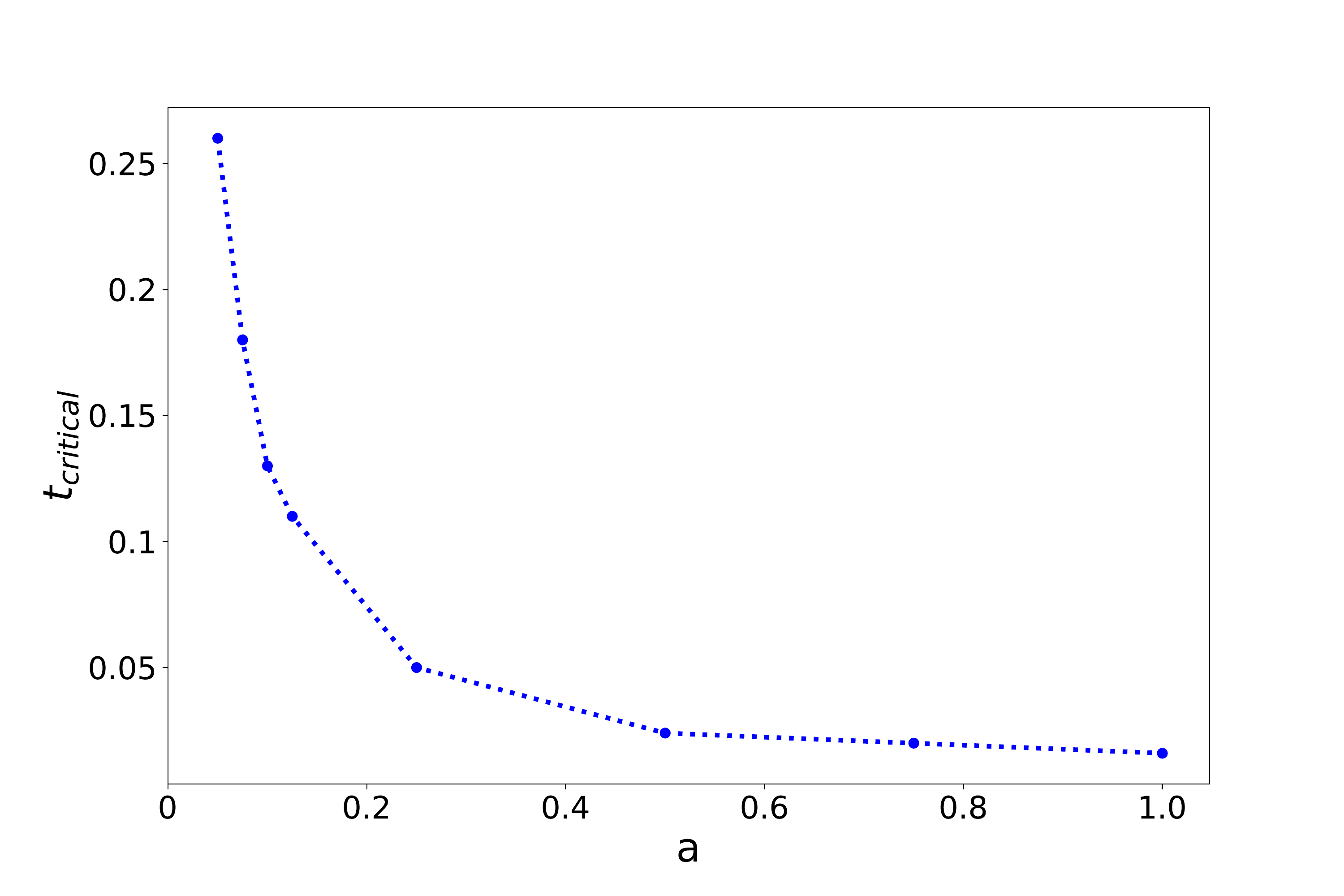}
\caption{Order parameter (left) and critical temperature (right) for different values of the scale factor. When the scale factor increases, the critical temperature decreases.}
\label{fig5}
\end{center}
\end{figure}
where $x \equiv \frac{ q }{b}$.  In Fig. (\ref{fig5}) , the order parameter $<x^2>$ for various scale factors $a(\tau)=1,0.1,0.05$ is plotted in the left panel, while the right panel demonstrates critical temperature versus scale factor, showing that the critical temperature decreases  as the scale factor increases. 

\section{Conclusions}

\label{sec5}

In this work, we have studied thermodynamical phase transition under MONDian gravity. In addition influence of the cosmic expansion on the critical temperature of the detected phase transition has been studied. We have shown that in the microcanonical ensemble of binary systems under MONDian gravity, a sharp phase transition is not present in the Newtonian gravity. Furthermore, we find a smoother phase transition with finite critical temperature by studying the specific heat  $C_v$ and an order parameter of a binary system in a canonical ensemble. One interesting result is that although both Newtonian and MONDian systems experience a phase transition in the canonical ensemble, they have different critical temperatures. The next steps in our research will be considering all interactions in the $N$-body system with the MONDian gravity and its connection with cosmological MOND \cite{2019ApJ...884L..25H}. In other directions, it would also be interesting to study the fractal structure of clusters in the crumpled phase of the system, as well as the equation of state (using the partition function in the
grand canonical ensemble) and complexity in the view of
\cite{2018PhRvD..97d4010H}. Investigation of the critical temperature of rotating MONDian self-gravitating systems will be another interesting problem.

\appendix

\section{ }
Here we calculate $g(E)$ for the binary system in MOND for the microcanonical ensemble. For the case with energy $-\frac{Gm^2}{b} <E< -\frac{Gm^2}{r_{M}}$ from Eq. (\ref{pv}), we have:
\begin{equation}
g(E)=\int_b^{-\frac{b {e_{2}}}{E}} r^2 (E+\frac{b {e_{2}}}{r})^2 \ dr=-\frac{b^3 (E+{e_{2}})^3}{3 E} \quad.
\end{equation} 
The temperature of this system with energy $E$ is given by $T^{-1} = \frac{\partial\ln(g(E))}{\partial E}$, 
\begin{equation}
t=\frac{\epsilon (\epsilon+1)}{2\epsilon-1}, 
\end{equation}
where dimensionless temperature and energy defined as $t=(bT/Gm^2)$ and $\epsilon =(bE/Gm^2)$ in Eq. (\ref{t}). If we increase the energy, the particles' accessible energy would be 
$\frac{-Gm^2}{r_{M}} <E<0$ making the integral (\ref{pv})
\begin{eqnarray*}
g(E) &= \int_b^{\frac{b {e_{2}}}{{e_{1}}}} r^2 \left(E+\frac{b {e_{2}}}{r}\right)^2 dr 
+\int_{\frac{b {e_{2}}}{{e_{1}}}}^{R e^ {\left(\frac{E}{{e_{1}}}\right)}} r^2 \left[E-{e_{1}} \log \left(\frac{r}{R}\right)\right]^2 dr \\
& = A_1 + A_2 +A_3,
\end{eqnarray*}
where $e_{1}=m\sqrt{Gm a_{0}}$, $e_{2}=Gm^2/b$ and 
\begin{eqnarray*}
A_1 &= \frac{\left(\frac{b E {e_{2}}}{{e_{1}}}+b {e_{2}}\right)^3-b^3 (E+{e_{2}})^3}{3 E} -\frac{b^3 {e_{2}}^3}{27 {e_{1}}^3}\left(6 E {e_{1}}+2 {e_{1}}^2\right),\\
A_2 &= \frac{1}{27} R^3 e^{\frac{3 E}{{e_{1}}}} [9 E^2+9 {e_{1}}^2 \log ^2\left(e^{E/{e_{1}}}\right)+ 6 E {e_{1}} -6 {e_{1}} (3 E+{e_{1}}) \log \left(e^{E/{e_{1}}}\right)+2 {e_{1}}^2],\\
A_3 &= -\frac{b^3 {e_{2}}^3}{27 {e_{1}}^3} \left[-6 {e_{1}} (3 E+{e_{1}}) \log \left(\frac{b {e_{2}}}{{e_{1}} R}\right)+9 {e_{1}}^2 \log ^2\left(\frac{b {e_{2}}}{{e_{1}} R}\right)+9 E^2\right].
\end{eqnarray*}

Similarly to the first part, we define the temperature of system in this range of energy as
\begin{equation*}
t = B_1 + B_2 + B_3, 
\end{equation*}
where 
\begin{eqnarray*}
B_1 &= b^3 \big[3 \log (\frac{b {e_{2}}}{{e_{1}} R}) (-3 \frac{e_{1}}{e_{2}} \log (\frac{b {e_{2}}}{{e_{1}} R})+6 \epsilon+2 \frac{e_{1}}{e_{2}})-9 \epsilon^2 ( \frac{e_{1}}{e_{2}})^2+3 \epsilon (7-9 \frac{e_{1}}{e_{2}}^2) +(\frac{e_{1}}{e_{2}}) (25-27 \frac{e_{1}}{e_{2}})\big],\\
B_2 &= (\frac{e_{1}}{e_{2}})^2 R^3 e^{\frac{3\epsilon {e_{2}} }{{e_{1}}}} (9 \epsilon^2+6 \epsilon\frac{e_{1}}{e_{2}}+3 \frac{e_{1}}{e_{2}} \log (e^{\epsilon {e_{2} }/{e_{1}}}) (3 \frac{e_{1}}{e_{2}} \log (e^{\epsilon {e_{2} }/{e_{1}}})-2 (3 \epsilon+\frac{e_{1}}{e_{2}})), \\
B_3 &= 2 (\frac{e_{1}}{e_{2}})^2
\big[3\frac{e_{1}}{e_{2}} R^3 e^{\frac{3 \epsilon}{{e_{1}}}} (9 e^2+6 \epsilon \frac{e_{1}}{e_{2}}+3 \frac{e_{1}}{e_{2}} \log (e^{\epsilon {e_{2} }/{e_{1}}}) (3 \frac{e_{1}}{e_{2}} \log (e^{\epsilon {e_{2} }/{e_{1}}})\\
&-2 (3 \epsilon+\frac{e_{1}}{e_{2}}))+2 (\frac{e_{1}}{e_{2}})^2-3 b^3 (-6 \log (\frac{b {e_{2}}}{{e_{1}} R})+6 \epsilon(\frac{e_{1}}{e_{2}})^2+9 (\frac{e_{1}}{e_{2}})^2 -7 ))^{-1}\big].
\end{eqnarray*}

Finally for the energy range $0<E<\infty $, we have:
\begin{equation*}
g(E)=\int_b^{\frac{b {e_{2}}}{{e_{1}}}} r^2 \big[E+\frac{b {e_{2}}}{r}\big]^2 dr +	\int_{\frac{b e_{2}}{e_{1}}}^R r^2 \big[E-{e_{1}} \log \left(\frac{r}{R}\right)\big]^2 dr = C_1 + C_2,
\end{equation*}
where
\begin{eqnarray*}
C_1 &= \frac{\left(9 E^2+6 E e_{1}+2 e_{1}^2\right) \left(e_{1}^3 R^3-b^3 e_{2}^3\right)}{27e_{1}^3}+\frac{3 b^3 e_{1} e_{2}^3 \log \left(\frac{b e_{2}}{e_{1} R}\right) \left[-3 e_{1} \log \left(\frac{b e_{2}}{e_{1}R}\right)\right]}{27e_{1}^3},\\
C_2 &= \frac{\left(6 E+2 e_{1}\right)}{27 e_{1}^3}+\frac{\left(\frac{b E e_{2}}{e_{1}}+a e_{2}\right)^3-b^3 (E+e_{2})^3}{3 E}.
\end{eqnarray*}
The temperature of the system is similar to parts one and two above, and is given by:
\begin{equation*}
t = D_1+ D_2+D_3,
\end{equation*}
where
\begin{equation*}
D_1 = \frac{b^3 \left[9 \epsilon^2(\frac{e_{1}}{e_{2}})^2+3 \epsilon \left(9(\frac{e_{1}}{e_{2}})^2 e_{2}-7 \right)+\frac{e_{1}}{e_{2}} (27\frac{e_{1}}{e_{2}}-25)\right]}{3 \left[b^3 \left(6 \epsilon(\frac{e_{1}}{e_{2}})^2+9(\frac{e_{1}}{e_{2}})^2-7 \right)-6 b^3 \log \left(\frac{b e_{2}}{e_{1} R}\right)-2(\frac{e_{1}}{e_{2}})^2 R^3 (3 \epsilon+e_{1})\right]},
\end{equation*}

\begin{equation*}
D_2 = \frac{3 b^3 e_{2}^3 \log \left(\frac{b e_{2}}{e_{1} R}\right) \left[3\frac{e_{1}}{e_{2}} \log \left(\frac{b e_{2}}{e_{1} R}\right)-2 (3 \epsilon+e_{1})\right]}{3 \left[b^3 \left(6 \epsilon(\frac{e_{1}}{e_{2}})^2+9(\frac{e_{1}}{e_{2}})^2 -7 \right)-6 b^3 \log \left(\frac{b e_{2}}{e_{1} R}\right)-2(\frac{e_{1}}{e_{2}})^2 R^3 (3 \epsilon+e_{1})\right]},
\end{equation*}

\begin{equation*}
D_3 = -\frac{e_{1}^2 R^3 \left[9 \epsilon^2+6 \epsilon\frac{e_{1}}{e_{2}}+2(\frac{e_{1}}{e_{2}})^2\right]}{3 \left[b^3 \left(6 \epsilon(\frac{e_{1}}{e_{2}})^2+9(\frac{e_{1}}{e_{2}})^2 -7 \right)-6 b^3 \log \left(\frac{b e_{2}}{e_{1} R}\right)-2(\frac{e_{1}}{e_{2}})^2 R^3 (3 \epsilon+e_{1})\right]}.
\end{equation*}

\end{document}